\documentstyle[prl,aps,multicol,psfig]{revtex}
\begin{document}

\draft

\title{Growth and Collapse of a Bose Condensate with Attractive Interactions}

\author{C.A. Sackett$^1$, H.T.C. Stoof$^{1,2}$, and R.G. Hulet$^1$}      
\address{$^1$Physics Department and Rice Quantum Institute,
             Rice University, Houston, Texas 77251-1892} 
\address{$^2$Institute for Theoretical Physics,
             University of Utrecht, Princetonplein 5, 
             3584 CC Utrecht, The Netherlands}
\maketitle

\begin{abstract}
We consider the dynamics of a quantum degenerate trapped gas of $^7$Li atoms.
Because the atoms have a negative $s$-wave scattering
length, a Bose condensate of $^7$Li becomes mechanically unstable when
the number of condensate atoms approaches a maximum value.  We calculate
the dynamics of the collapse that occurs when the unstable point is reached.  
In addition, we use the quantum Boltzmann equation to investigate the
nonequilibrium kinetics of the atomic distribution during and after
evaporative cooling.  The condensate is found to
undergo many cycles of growth and collapse before a stationary state is
reached.
\end{abstract}

\pacs{PACS number(s): 03.75.Fi, 67.40.-w, 32.80.Pj, 42.50.Vk}

\begin{multicols}{2}

The recent observations of Bose-Einstein 
condensation in weakly interacting gases \cite{BEC1}
has enabled a series of beautiful experiments that 
probe the dynamics of the condensate.  The
frequency and damping of the collective modes 
of a condensate \cite{BEC2}, propagation of sound in a
condensate \cite{MIT2}, and recently, the growth of 
the condensate \cite{wolfgang}, have been reported.  Although
these experiments have used atoms with positive 
$s$-wave scattering lengths, we show in this
paper that the dynamical behavior of a negative 
scattering length gas, such as $^7$Li, is especially
interesting, and offers the opportunity to directly observe 
and study macroscopic quantum tunneling.

A negative scattering length $a$ implies effectively attractive 
interactions.  In a spatially homogeneous gas, these interactions lead to 
ordinary classical condensation into a liquid or
solid, preventing Bose condensation in the metastable region 
of the phase diagram \cite{henk1}. However, confinement in an atom 
trap produces stabilizing forces that enable the formation of a
metastable Bose condensate, if the number of condensed atoms is less than 
some maximum number $N_m$.  For a harmonic trapping potential, 
Ruprecht {\it et al.} \cite{keith1} showed that in mean-field
theory $N_m \simeq  0.57 l/|a|$, where $l = (\hbar/m \omega )^{1/2}$ is 
the extent of the one-particle ground state in the
harmonic trap \cite{marianne}.  For the $^7$Li experiments of Ref. \cite{RICE}, 
$N_m \simeq  1400$ atoms, which agrees with the measured value.

Although a condensate can exist in a trapped gas, it
is predicted to be metastable and to decay by quantum or thermal
fluctuations \cite{kagan,shuryak,henk2}.  The condensate has only one 
unstable collective mode, which in the case of an isotropic trap 
corresponds to the breathing mode \cite{keith2,dan}. The condensate therefore
collapses as a whole, either by thermal excitation over a macroscopic energy
barrier, or by quantum mechanical tunneling through it.
The probability of forming small, dense clusters is
greatly suppressed because of the large energy barrier for this 
process, compared to that for the breathing mode.  This suppression can also 
be understood from the fact that the typical length scale for fluctuations of 
the condensate is the healing length, which is precisely of the order of 
the condensate size near the instability point.

Experimentally, it is also important
to understand how such a condensate can be formed from a noncondensed thermal 
cloud by means of evaporative cooling. This question was recently addressed by 
Gardiner {\it et al.} in the context of the experiments with gases
having $a>0$ \cite{peter1}.  These authors neglect the 
coherent dynamics of the condensate and focus instead on the kinetics of 
condensation \cite{henk3}. By treating the noncondensed atoms as a 
static `heat bath' with a chemical potential that is larger than the ground 
state energy, they are able to derive a simple equation for the growth of the 
number of condensate particles that appears to fit well with 
experimental results \cite{wolfgang}. The same approach, however, does not work 
in the case of atomic $^7$Li because it does not allow for the collapse of the 
condensate.  Moreover, 
the process of evaporative cooling leads to dynamical changes in the 
noncondensed `heat bath'. The study of both these effects on the 
dynamics of the 
condensate is the main topic of this Letter. Some preliminary results 
have already appeared in a recent review article\cite{cass}.

When collisions can be neglected, 
the dynamics of the condensate wavefunction $\psi$
for a gas with $a<0$ is well-described by
the nonlinear Schr\"odinger (or Gross-Pitaevskii \cite{GP}) equation
\begin{eqnarray}
\label{NLSE}
i\hbar \frac{\partial \psi({\bf x},t)}{\partial t} =
  \left( - \frac{\hbar^2 \nabla^2}{2m} + V({\bf x}) 
                           \right) \psi({\bf x},t) \hspace*{0.4in} \nonumber \\
    + T^{2B}({\bf 0},{\bf 0};0) |\psi({\bf x},t)|^2 \psi({\bf x},t)~.
\end{eqnarray}
Here $m$ denotes the mass of $^7$Li, 
$T^{2B}({\bf 0},{\bf 0};0) = 4\pi a \hbar^2/m$ is the two-body T-matrix, 
and $a$ is the scattering length of $-27\,a_0$ \cite{randy}. For the 
external trapping potential $V({\bf x})$ we take a harmonic potential with an 
effective isotropic level splitting 
$\hbar\omega = \hbar(\omega_x\omega_y\omega_z)^{1/3}$ of 
7 nK \cite{RICE}. 
Note that we ignore the mean-field contribution from 
the noncondensed atoms in this equation \cite{tony}, because 
it is nearly constant over the size of the condensate and
therefore only slightly affects the condensate dynamics.

This description is semiclassical, and also neglects quantum and thermal 
fluctuations. These fluctuations are most easily incorporated by means of the 
partition function of the condensate, which is a functional integral 
$\int d[\psi^*]d[\psi]~\exp(-S[\psi^*,\psi]/\hbar)$ over all periodic 
configurations of the condensate, with a weight determined by the (Euclidean) 
action of the nonlinear Schr\"odinger equation. It is 
most convenient to calculate this partition function in terms of 
the density and phase of the condensate, defined by 
$\psi = \sqrt{\rho} e^{i\chi}$. The Gaussian integral over the phase field 
$\chi$ can then be performed exactly leaving only the determination of 
the functional integral $\int d[\rho]~\exp(-S[\rho]/\hbar)$. 

Unfortunately, this integral cannot be calculated exactly. 
However, since we are 
primarily interested in the dynamics of the unstable breathing mode of the 
condensate, we can use a variational method \cite{henk2,peter2} and 
consider only Gaussian density profiles
\begin{eqnarray} \label{gaussden}
\rho({\bf x};q(t)) 
   = N_0 \left( \frac{1}{\pi q^2(t)} \right)^{3/2} 
         \exp \left( - \frac{ {\bf x}^2 }{q^2(t)} \right)~.
\end{eqnarray}
Substituting this profile in the action $S[\rho]$, we find that the dynamics of 
the condensate is equivalent to the dynamics of a fictitious particle with   
mass $m^* = 3N_0 m/2$ in the unstable potential \cite{better}
\begin{equation}
U(q) = N_0 \left( \frac{3\hbar^2}{4mq^2} + \frac{3}{4} m\omega^2 q^2 
                  - \frac{N_0}{\sqrt{2\pi}} \frac{\hbar^2 |a|}{mq^3} \right)~.
\end{equation}  

\begin{figure}
\psfig{figure=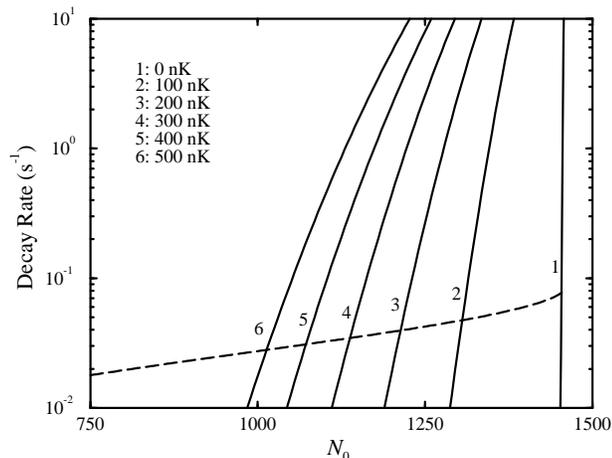}
\caption{\protect \narrowtext
         Decay rate of the condensate as a function of the number of condensate
         particles at $0$, $100$, $200$, $300$, $400$, and $500 nK$. The 
         dashed line shows the decay due to inelastic two and three-body
         collisions \protect \cite{Moerdijk96}. 
         \protect \label{rates}}
\end{figure}

The rate of decay for both quantum tunneling and thermal fluctuations can now
easily be calculated within this formalism \cite{henk2} and are shown in 
Fig.~\ref{rates}.
For large numbers of condensate atoms,
these collective decay mechanisms are much faster than the 
decay caused by inelastic two and three-body collisions,
since the energy barrier out of the metastable 
minimum vanishes as $N_0$ approaches $N_m$.  For 
experimentally relevant temperatures of 300 to 500 nK, the collective
decay processes dominate for $N_0$ greater than about 1100 atoms.

\begin{figure}
\psfig{figure=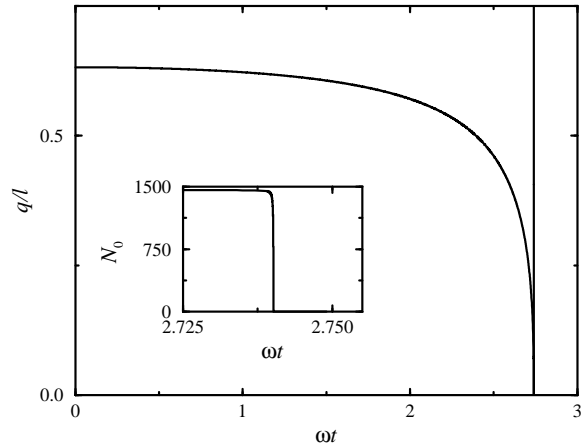}
\caption{\protect \narrowtext
         Typical evolution of the condensate during collapse.
         \protect \label{dynamics}}
\end{figure}

Besides the calculation of the decay rates, the above anology also allows 
study of the dynamics 
of the collapse that occurs after the condensate has been driven 
out of its metastable minimum. A typical example is shown in 
Fig.~\ref{dynamics}. From this figure we see that the condensate first 
collapses with increasing speed along the potential hill outside the barrier. 
However, during the collapse, the condensate density grows rapidly,
thereby increasing the decay rate for inelastic two and three-body collisions. 
Atoms that inelastically collide acquire subtantial energy and
are ejected from the trap.
Because of these loss mechanisms, the collapse is arrested when
the number of condensate atoms is of the order of one.
Atoms are lost so quickly that the density of the condensate always obeys 
$n|a|^3 \ll 1$, which is a necessary requirement for the validity of the
nonlinear Schr\"odinger equation (\ref{NLSE}).
If there were no inelastic collisions,
the condensate would fully collapse \cite{lev}, and a more complex
theory would be needed.

The above remarks pertain to the dynamics of the condensate in the absence
of collisions with noncondensed atoms, but so far all experiments producing 
BEC have relied on evaporative cooling, which requires such collisions.  
Therefore, to investigate the kinetics of condensation in a trapped gas,
we use the Boltzmann transport equation in a 
way similar to the treatment of evaporative cooling in 
Refs.~\cite{Luiten96,Sackett97}. In this approach, we define a semiclassical 
distribution function $f({\bf x},{\bf p})$,
with the number of atoms at position ${\bf x}$ with momentum ${\bf p}$ being
$d{\bf x} d{\bf p} \, f({\bf x},{\bf p})/(2 \pi \hbar)^{3}$. The 
evolution of the distribution function is then given by \cite{Huang87}
\begin{equation}
\label{eq:boltz}
	\frac{d}{dt} f({\bf x},{\bf p},t) = I({\bf x},{\bf p}) - 
		\Gamma({\bf x}) f({\bf x},{\bf p})~,
\end{equation}
where the derivative on the left is the total time derivative.
The effect of elastic collisions is given by
\begin{eqnarray}
\label{collint}
    I({\bf x},& {\bf p}_1 &)  =  \frac{\sigma}{4\pi^{4}m\hbar^{3}}\,
         \int d{\bf p}_2\,d{\bf p}_3\,d{\bf p}_4\, \nonumber \\
        & \times & \delta^3({\bf p}_{1}+{\bf p}_{2}-{\bf p}_{3}-{\bf p}_{4})
        \,\delta(p_{1}^{2}+p_{2}^{2}-p_{3}^{2}-p_{4}^{2}) \nonumber \\
        & \times & [ f_3 f_4 (1+f_1)(1+f_2) - f_1 f_2 (1+f_3)(1+f_4)],
\end{eqnarray}
where the factors $(1+f_i)$ account for the Bose statistics of the atoms.
Here $f_i$ stands for $f({\bf x},{\bf p}_i)$ and $\sigma = 8\pi a^2$.
Inelastic collisions, which lead to a loss of atoms from the trap, are
described by 
$\Gamma({\bf x}) = G_{1} +
        G_{2} \int d{\bf p} \, f({\bf x},{\bf p})/(2 \pi \hbar)^{3}$,
where the terms with $G_1$ and $G_2$
reflect the loss of atoms due, respectively,
to collisions with hot background gas atoms and to inelastic
collisions between trapped atoms.  

As for the case of a classical gas \cite{Luiten96,Sackett97}, 
Eq.~(\ref{eq:boltz}) can be simplified
by taking the motion of the atoms to be ergodic, so that
an atom with a given energy will sample each available cell in
$\{{\bf x},{\bf p}\}$ phase space with equal probability.  
The distribution function $f({\bf x},{\bf p})$ 
can then be expressed in terms
of the energy distribution function $f(E)$, through the relation
\begin{equation} \label{eq:ergodic}
 f({\bf x},{\bf p}) = \int dE\,\delta(H({\bf x},{\bf p})-E)\,f(E)\,
\end{equation}
where $H({\bf x},{\bf p}) = p^2/2m + V({\bf x})$ is the classical Hamiltonian.
The differential equation for $f(E,t)$ is derived by substituting
(\ref{eq:ergodic}) into (\ref{eq:boltz}), as is described in detail
in Refs.~\cite{Luiten96,Sackett97}.  
The only difference here is the use of Bose statistics, which requires the 
insertion of factors $1+f(E_i)$ as in Eq.~(\ref{collint}).
Although the semiclassical approximation results in a continuous
function $f(E)$, the quantum nature of the trapped gas can be recovered
by restricting $f(E)$ to values $E_n = (n + 3/2) \hbar \omega$.

The above model neglects the effect of the mean-field interaction energy.
Per atom, however, this interaction energy is limited to about
$\hbar \omega$ by the stability criterion 
arising from $a$ being negative, which limits the effect of the mean-field
energy on the kinetics of the gas.
The accuracy of the approximations can be checked by considering collisions
between condensate atoms, since the mean-field interaction is largest in the 
condensate and
the semiclassical approximation is least accurate for the ground state of
the trap.  We compare the rates for inelastic collisions
between condensate atoms, $G_2 \int d{\bf x} \rho({\bf x})^2$, which
depend only on the density.  
The exact condensate density is determined from Eq.~(\ref{NLSE}), using
the variational method of Eq.~(\ref{gaussden}).  In contrast, 
the semiclassical density derived from Eq.~(\ref{eq:ergodic}) is
\begin{equation}
	\rho({\bf x}) = \left ( \frac{2}{3\pi} \right )^2 
		\left ( \frac{N_0}{l^3} \right ) 
		\left ( 3-\frac{x^2}{l^2} \right )^{1/2}.
\end{equation}

From these distributions, the ratio of the exact and approximate collision
rates is $1.2 (l/q)^3$.  When interactions are
small, then $q \simeq l$, and the error caused by the semiclassical 
approximation is only 20\%.  However, the mean-field interaction causes $q$ to 
decrease as $N_0$ is allowed to grow. From the variational calculation we find 
that $q = 0.67 l$ just before the collapse. The error in the collision
rate is then a factor of 4, and collision terms for other low-lying states will 
be incorrect by similar, though smaller, amounts. Although these errors are 
significant, they only occur for $N_0$ relatively close to $N_m$. For instance 
at $N_0 \simeq 0.8\,N_m$, the error is a factor of 2.  As will be seen below,
the condensate number is greater than $0.8\,N_m$ for only a small fraction of 
time, so the approximations are useful, but for $N_0 \simeq N_m$ the 
quantitative predictions of the model will be somewhat inaccurate.

The possibility for the condensate to collapse is included in the model
using the decay rates given in Fig.~\ref{rates}. A random number is 
chosen to determine whether a decay occurs during an integration time
step, and when one does the condensate number is set to zero reflecting the 
rapid loss shown in Fig.~\ref{dynamics}. Also, evaporative cooling is included 
by setting $f(E) = 0$ for $E > E_c(t)$, where $E_c(t)$ is the experimentally 
settable cutoff energy.

\begin{figure}
\psfig{figure=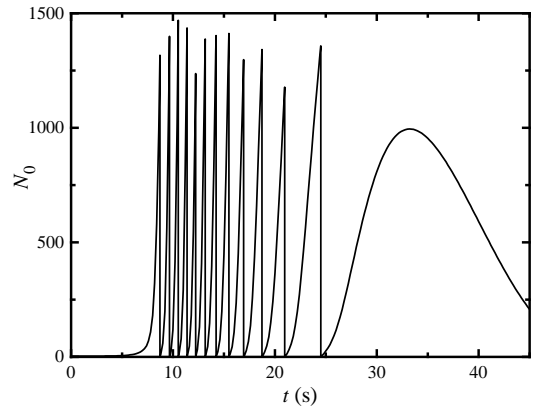}
\caption{\protect \narrowtext
         Typical evolution of condensate number $N_0$ in response to evaporative
         cooling. The gas initially consists of $4 \times 10^5$ atoms at 500 nK. 
         During the first 5 seconds of evolution, roughly 40\% of the atoms are
         removed by evaporative cooling, after which evaporation is halted.
         \protect \label{fig:N0}}
\end{figure}

The response of the condensate number to evaporative cooling
is shown in Fig.~\ref{fig:N0}.  In this case, $E_c(t)$ consists of a 5-second 
period of rapid cooling, after which evaporation is halted and the gas
allowed to equilibrate. The subsequent behavior shows the condensate
alternately filling and collapsing, until finally the phase space density
of the gas is too low to reach $N_0 \simeq N_m$. The slow final decay is
due to the trap losses.  

\begin{figure}
\psfig{figure=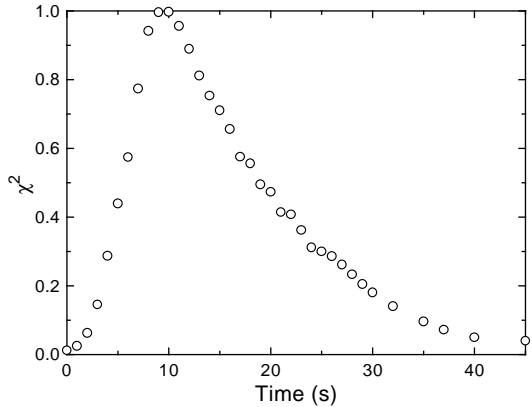}
\caption{\protect \narrowtext
         Equilibration of degenerate gas. Using the same conditions as in 
         Fig.~\protect \ref{fig:N0}, at each time the density distribution is   
         fit to an equilibrium density distribution, and the resulting value of 
         $\chi^2$ plotted. 
	 \protect \label{fig:chi2}}
\end{figure}

As shown in the figure, the time for which the oscillations in $N_0$ persist is 
anomalously long compared to the elastic collision
time $\tau_c = 1/\langle n \sigma  v \rangle \simeq 0.8$ s.
In order to investigate this phenomenon further, 
the density profiles generated by the model were fit to equilibrium 
Bose-Einstein distributions at various times.  
The values of $\chi^2$ resulting from the fits are 
plotted in Fig.~\ref{fig:chi2}.
After reaching degeneracy, the curve shows an approximately
exponential approach to equilibrium, with a time constant of 
about $10\,\tau_c$.
In contrast, the same test performed on a nondegenerate cloud 
initially prepared in a nonequilibrium state yields an equilibration time
of 5$\,\tau_c$.  The relatively slow approach to equilibrium for the
degenerate case is presumably caused by the small phase-space volume of the 
condensate and the fact that due to the limit $N_0 < N_m$,
the stimulated Bose scattering factors in Eq.~(\ref{eq:boltz}) cannot
become as enormous as they might in the $a>0$ case. 
Furthermore, the oscillations in $N_0$ will
persist until the distribution is very close to equilibrium, since the
total number of atoms in the trap is much greater than $N_m$.

Note, finally, that the evolution in Fig.~\ref{fig:N0} represents only one 
possible outcome of evaporative cooling, and that because of the stochastic 
nature of the collapse a given evolution is not repeatable. Experimental
measurements of $N_0$ in the degenerate regime should, however,
exhibit large fluctuations between 0 and $N_m$. Observation of such fluctuations 
and measurement of their statistics would provide confirmation of the collective 
nature of the collapse. If sufficiently low temperatures can be reached,
these fluctuations would be evidence of macroscopic quantum tunneling.
We therefore believe that a dilute Bose gas with $a<0$ 
presents unique opportunities for studying the dynamical properties of a 
condensate.

\end{multicols}

\begin{references}
\bibitem{better} If the density field is expanded as
                 \begin{eqnarray}
                 \rho({\bf x};q(t)) + 
                   {\sum_{n \ell m}}' C_{n \ell m}(t) e^{-{\bf x}^2/2q^2(t)}
                                \phi_{n \ell m}({\bf x};q(t))~,  \nonumber
                 \end{eqnarray}
                 where $\phi_{n \ell m}({\bf x};q)$ are the usual harmonic
                 oscillator states with frequency $\omega_q = \hbar/mq^2$, and
                 the sum does not include the states $\phi_{000}$ and
                 $\phi_{100}$, we can systematically improve upon this 
                 result by expanding the action $S[\rho]$ up to 
                 quadratic order in $C_{n \ell m}$ and integrating out
                 the coefficients describing the dynamics of all the 
                 stable collective modes.
\end{references}
\end{document}